\documentclass[a4paper,11pt]{article}
\pdfoutput=1 

\usepackage{jcappub} 

\usepackage[T1]{fontenc} 

\title{Growth of Linear Perturbations in a Universe with Superfluid Dark Matter}

\author[a,1]{Shreya Banerjee \note{Corresponding author}}
\author[b]{Sayantani Bera}
\author[c]{and David F. Mota}

\affiliation[a]{Department of Physics, Ben-Gurion University, P.O.Box 653, Beer-Sheva 84105 Israel}
\affiliation[b]{Inter-University Center for Astronomy and Astrophysics, Post Bag 4, Ganeshkhind, Pune-411007, India}
\affiliation[c]{Institute of Theoretical Astrophysics, University of Oslo, P.O. Box 1029 Blindern, N-0315
Oslo, Norway}

\emailAdd{banerjee@post.bgu.ac.il}
\emailAdd{sayantani@iucaa.in}
\emailAdd{d.f.mota@astro.uio.no}

\abstract{The Lambda-Cold Dark Matter (LCDM) model agrees with most of the cosmological observations, but has some hindrances from observed data at smaller scales such as galaxies. Recently, Khoury and Berezhiani proposed a new theory involving interacting superfluid dark matter with three model parameters in \cite{khoury2015}, which explains galactic dynamics with great accuracy. In the present work, we study the cosmological behaviour of this model in the linear regime of cosmological perturbations. In particular, we compute both analytically and numerically the matter linear growth factor and obtain new bounds for the model parameters which are significantly stronger than previously found. These new constraints come from the fact that structures within the superfluid dark matter framework grow quicker than in LCDM, and quite rapidly when the DM-baryon interactions are strong.}


\begin{document}
\maketitle
\flushbottom

\section{Introduction}
With the advent of precision cosmology and satellites like Planck and WMAP, we have gained new insights about the evolution of the universe. Till date, Lambda-Cold Dark Matter (LCDM) provides the best fit to these available data and has been widely accepted as the standard model of cosmology \cite{Planck}. The hypothesis of CDM, which are assumed to be collisionless non-relativistic particles, along with baryonic matter explains the CMB temperature anisotropy, matter power spectra, large scale galaxy distributions and lensing data remarkably well. In fact, the abundance of galaxy clusters and observed large scale structure formation history strongly supports the collisionless CDM scenario as opposed to any alternative theories to LCDM \cite{ref1,ref2,ref3}. However, at smaller scales, CDM faces a number of challenges that need to be addressed \cite{lcdm1}. For example, the Baryonic Tully-Fisher relation and the corresponding tight correlation between the mass and dispersion velocity at the high-mass end can not satisfactorily be explained by CDM halo which predicts a larger scatter due to feedback processes in the galaxy \cite{lcdm1.5}. Apart from this, there is another issue with the standard CDM picture in the galactic scale, known as the {\it cusp-core problem} \cite{lcdm2}. The simulations of galactic halos with CDM produce a kink (cusp) at the center of the galaxy, whereas observations of various galactic density profile suggest a flat core. With improved observations of the faint dwarf galaxies and substructures within the galaxies like Milky Way and Andromeda, new set of discrepancies arise. While the missing satellite problem in dwarf galaxies (\cite{lcdm3}) has been addressed to some extent, the {\it To Big To Fail Problem}, arising from the prediction of satellites that are too massive and too dense by LCDM, compared to those observed, still remains unresolved \cite{lcdm4,lcdm5}.

\noindent Due to the above unresolved issues, scientists have looked into other alternative explanations through modifications of General Relativity (GR). Several models have been proposed so far with the aim to explain existing data to the same degree of accuracy as LCDM as well as overcome its drawbacks. Many of them have already been ruled out or are highly constrained by the ongoing observations of gravitational waves, but some theories like $f(R)$, $f(T)$, $f(\mathcal{G})$, Scalar-tensor-vector theories of gravity etc. are still consistent with the data, and new observations are required to falsify these theories \cite{MG1,MG2, MG3, MG4,MG5,MG6,MG7}. These theories are relativistic corrections of GR which modify the dynamics of spacetime through the modified field equations. The theory of Modified Newtonian Dynamics (MOND), on the other hand, is a modification to the Newtonian force law that changes the dynamics of interaction between two massive bodies in the non-relativistic limit \cite{MOND1,MOND2}. MOND was first proposed in 1983 by Milgrom to account for the flattened galaxy rotation curves near the edge of the spiral galaxies like Milky Way. There is a universal acceleration scale $a_0$ in MOND, whose value is obtained as $10^{-8} cm/s^2$. For accelerations much lower than this scale, the Newtonian law is modified, and this explains the flat galaxy rotation curve data for a large number of galaxies \cite{MOND3}. Interestingly, the Baryonic Tully-Fisher relation in galaxies can exactly be derived from MOND where $M \propto v_c^4$. MOND can also explain several other galactic observations like the planar structure of galaxies, low merger rate etc \cite{MOND4}. Thus, we see that MOND, with just one free parameter, is a very well-behaved theory at the galactic scale. However, despite these successes, MOND faces several challenges in extragalactic and cosmological scales. Proper relativistic extension of MOND is not available \cite{MOND5}. Hence it cannot be applied at cosmological scales.

\noindent The effectiveness of MOND at small scales and success of LCDM at cosmological scales are the main motivations for scientists to look for models which are CDM-MOND hybrids i.e., theories that include usual cold dark matter as collisionless particles at cosmological scales, but give rise to a MOND-like modified force law at galactic scales such that they satisfy both sets of observations. This class of models take into consideration the interacting dark matter-baryon picture where a MOND-like force is mediated  through this new interaction term. Based on this idea, many models have been proposed which can reproduce both CDM features as well as MOND in their respective regime of validity \cite{IDM1,IDM2,IDM3,khoury2017.2,khoury2015}.

\noindent In this paper we shall focus on one such model proposed recently by Khoury and Berezhiani \cite{khoury2015}, where CDM can form condensates at galactic scales depending upon the surrounding temperature and can behave as superfluid. It has already been shown by the authors that such model can explain a number of galactic scale observations due to their MONDian behaviour, which normal CDM fails to explain \cite{khoury2015,khoury2017,khoury2016,khoury2016.2}. There are two free parameters in the theory which is assumed to be temperature dependent. It has been argued that at cosmological scales, the theory behaves as usual CDM and thus the background evolution and other cosmic histories remain unchanged as compared to LCDM. Here, we study the cosmological evolution of the background as well as the matter perturbations. We check whether the present model remains well-behaved at cosmological scales as has been claimed by the authors and compare our results with LCDM.

\section{Dark Matter Superfluid-Overview}
The central idea of this model is that CDM is made up of particles which undergoe phase transition  below a particular critical temperature and becomes a superfluid. This requires that the particle CDM needs to be strongly interacting below a particular temperature. The superfluid behaviour depends on the strength of interaction and the mass of the particle. It has been shown in \cite{khoury2015} that in order to form a Bose-Einstein Condensate (BEC) the following condition must be satisfied
\begin{equation}
m \lesssim \left(\frac{\rho}{v^3}\right)^{1/4}
\end{equation}
where $m$ and $v$ corresponds to the mass and velocity of the particle respectively and $\rho$ is the density of the condensate. Assuming virialization of dark matter halo at galactic scales, this gives an upper bound on the mass of the particle forming the halo
\begin{equation}
m \lesssim 2.3(1+z_{vir})^{3/8}\left(\frac{M}{10^{12}h^{-1}M_{\bigodot}}\right)^{-1/4} {\rm eV}
\end{equation}
Further assuming thermalization of CDM particles, one obtains the bound on interaction cross section as
\begin{equation}
\frac{\sigma}{m}\gtrsim 52(1+z_{vir})^{-7/2}\left(\frac{m}{eV}\right)^4\left(\frac{M}{10^{12}h^{-1}M_{\bigodot}}\right)^{2/3} {\rm cm^2g^{-1}}
\end{equation}
Using equipartition law, the critical temperature $T_c$ of the CDM condensate can be obtained as
\begin{equation}
T_c=6.5\left(\frac{\rm eV}{\rm m}\right)^{5/3}(1+z_{vir})^2 {\rm mK}
\end{equation}
It has been argued in \cite{khoury2015} that the temperature of CDM at cosmological scales is much below the critical temperature (${\mathcal{O}}(10^{-28}$) for $m\sim$ eV) which implies that the condensate behaves as a $T\approx 0$ superfluid at cosmological scales.\\

The description of superfluid dark matter is given in terms of a low energy effective theory with the Lagrangian of the form:
\begin{equation}
\mathcal{L} = \frac{2\Lambda(2m)^{3/2}}{3}\left(\dot{\theta}-m\Phi-\frac{(\nabla\theta)^2}{2m}\right)^{3/2}
\end{equation}
Let us now understand the motivation of choosing such a Lagrangian. Here, $\theta$ is the phase of the wavefunction describing the superfluid phonon modes and $\Phi$ is the gravitational potential in which the DM particle sits and is given by the standard Newtonian potential in the usual non-relativistic case. This Lagrangian has a free parameter $\Lambda$ which defines the strength of the superfluid (i.e. defined by the number of particles in the condensate state). The power of the Lagrangian is defined by the choice of the equation of state (EoS), and a fractional power of $5/2$ is indeed obtained in superfluids formed by ultra cold atoms. In the case of CDM superfluid, the choice of the power $3/2$ in the Lagrangian is somewhat arbitrary, but motivated by the fact that the superfluid DM should give rise to MOND-like dynamics at galactic scales when baryons are also included. This also corresponds to an equation of state $P \sim \rho^3$ which is suggestive of a dominant three-body interaction process. What kind of particles can lead to such a superfluid with this particular EoS and the physics of its formation has not been discussed earlier and is beyond the scope of this paper. For our purpose, we shall assume the form of this Lagrangian to study the characteristic features of the resultant superfluid DM model.\\

In the effective field theory formalism, the superfluid is described in terms of interacting phonon modes. The phonon modes can be described by the scalar field $\theta$, which, at a constant chemical potential $\mu$, can be expanded as,
$$ \theta = \mu t+ \phi $$ where $\phi$ denotes the excitation of the phonon modes.\\

The DM superfluid couples to the baryons through the phonon modes via an interaction given by the Lagrangian:
\begin{equation}
\mathcal{L}_{int}= -\alpha\frac{\Lambda}{M_{Pl}}\theta \rho_b
\end{equation}
This kind of interaction ensures a MOND force. Here $\alpha$ is a dimensionless free parameter, which sets the interaction strength of the interaction, and $\rho_b$ is the baryonic mass density.\\

Thus, the complete Lagrangian for an interacting superfluid DM is given by,
\begin{equation}
\mathcal{L}=\frac{2\Lambda(2m)^{3/2}}{3}\left(\dot{\theta}-m\Phi-\frac{(\nabla\theta)^2}{2m}\right)^{3/2}-\alpha\frac{\Lambda}{M_{Pl}}\theta \rho_b
\label{sflagrangian}
\end{equation}
It has been shown in \cite{khoury2015} that the MONDian acceleration arises as a special case of the dynamics of the above Lagrangian. The validity of this model in solar system and Bullet cluster has also been discussed there.\\

In the cosmological context, although the authors in \cite{khoury2015} discuss some general points regarding the background behaviour and the equation of state of this new superfluid dark matter, they do not shed much light on other important points such as growth of perturbations and structure formation. In the next sections, we solely focus on the cosmological aspects of this new theory. 
\section{Cosmological Solutions}
In this section, we will study this theory in cosmological context. This is of particular interest since the theory also needs to be consistent with the present cosmological data. 
\subsection{Background Solutions}
For the background cosmology, we have $\theta= \theta(t)$. In the FLRW background with a scale factor $a$, the equation of motion for $\theta$ can be derived from the action as,
\begin{equation}
\frac{d}{dt}\left[(2m)^{3/2}a^3 \dot{\theta}^{1/2}\right]= -\frac{\alpha}{M_{Pl}}a^3\rho_b
\end{equation}
Assuming the evolution of baryons i.e. $\rho_b \propto 1/a^3$ as in standard LCDM, we get,
\begin{equation}
\rho_{DM} = -\frac{\alpha\Lambda}{M_{Pl}}m\rho_b t + \frac{m\Lambda C}{a^3}
\end{equation}
Here $C$ is an integration constant which has to be determined from the present DM density. The second term ($\rho_{dust}$) corresponds to the dust like evolution. For the second term to dominate (such that $\rho_{DM}$ behaves as dust), it can be shown that one needs to satisfy the following constraint:
\begin{equation}
\frac{\alpha\Lambda\rho_b}{M_{Pl}\rho_{dust}}mt_0 \leq 1
\label{cosmobound1}
\end{equation}
where $t_0$ is the present age of the universe. \\

{\it Bounds on the model parameters}:\\

\begin{itemize}
\item {\bf From the EoS}- Equation of state for the DM superfluid (assuming negligible interaction) is given by,
\begin{equation}
w=\frac{\rho_{dust}^2}{12 \Lambda^2 m^6}
\end{equation}
For DM to behave as dust at the background level, $\Lambda$ should be bounded from below,
\begin{equation}
\Lambda \gg 0.1 \left(\frac{m}{{\rm eV}}\right)^{-3} {\rm eV}
\label{cosmobound2}
\end{equation}
\item {\bf From coupling to baryons}- From \eqref{cosmobound1} and \eqref{cosmobound2}, and assuming a constant baryon-to-DM ration ($\rho_{dust}/\rho_b=6$), we get,
\begin{equation}
\alpha \ll 2.4 \times 10^{-4} \left( \frac{m}{{\rm eV}}\right)^2
\end{equation}
\end{itemize}

These bounds are different from the bounds obtained for galaxies, as discussed in \cite{khoury2015,khoury2017}.
\subsection{Perturbations}
Study of linear perturbation theory in the context of LCDM has been has been an important step towards understanding the evolution of the universe. CMB spectra carries information about the inhomogeneities present in the early universe. Hence, any cosmological model needs to satisfy the CMB data to a high degree of accuracy. This requires analysing the matter power spectrum resulting from the initial density perturbations. In this section, we examine the growth of cosmological perturbations in DM superfluid model at linear order.

The Lagrangian of the theory in an FLRW matter dominated universe is given as,
\begin{equation}
\mathcal{L}=c_1a^3\left(\dot{\theta}-\frac{(\nabla \theta)^2}{2m}-m\Phi \right)^{3/2}-c_2a^3\rho_b\theta
\label{lagrangian}
\end{equation}
where $c_1,c_2$ are constants expressed as,
\begin{eqnarray}
c_1 &=& \frac{2\Lambda(2m)^{3/2}}{3} \nonumber \\
c_2 &=& \alpha\frac{\Lambda}{M_{Pl}} 
\end{eqnarray}

Let us now perturb the phonon field and the baryon density around their background values such that,
$$\theta= \bar{\theta}+\delta \theta $$
$$\rho_b= \bar{\rho}_b+\delta \rho_b $$
In the weak field limit, the gravitational potential satisfies the Poisson equation:
\begin{equation}
\nabla^2 \Phi = \delta \rho_b + \delta \rho_{m}
\end{equation}
where we have assumed $4 \pi G=1$. The DM density is given by,
$$\rho_{m}=m \Lambda (2m)^{3/2}\dot{\theta}^{1/2}$$
Then perturbation is written in terms of the perturbations in the phonon field $\theta$ as,
$$\delta \rho_{m}=\frac{m}{2} \Lambda (2m)^{3/2}\frac{\delta \dot{\theta}}{\dot{\bar{\theta}}^{1/2}}$$
and thus we have,
$$\delta_{m}= \frac{\delta \rho_{m}}{\bar{\rho}_{m}}= \frac{\delta \dot{\theta}}{2\dot{\bar{\theta}}}$$

The Poisson equation can be written in momentum space as follows:
\begin{equation}
k^2 \Phi = \delta \rho_b + \delta \rho_{m}
\end{equation}
Considering the perturbations to be very small compared to the background values, we expand \eqref{lagrangian} upto leading order,
\begin{equation}
\mathcal{L}=c_1a^3\dot{\bar{\theta}}^{3/2}\left(1+\frac{3}{2}\frac{\delta \dot{\theta}}{\dot{\bar{\theta}}}-\frac{3}{4m}\frac{(\nabla \delta \theta)^2}{\dot{\bar{\theta}}}-\frac{3m}{2k^2}\frac{\delta \rho_b}{\dot{\bar{\theta}}}-\frac{3m^2\Lambda(2m)^{3/2}}{4k^2}\frac{\delta \dot{\theta}}{\dot{\bar{\theta}}^{3/2}}\right)-c_2a^3(\bar{\rho}_b+\delta \rho_b)(\bar{\theta}+\delta \theta)
\end{equation}
From the above equation we can get the equation of motion for $\bar{\theta}$.
\begin{equation}
\frac{d}{dt}\left[c_1a^3\left(\frac{3}{2}\dot{\bar{\theta}}^{1/2}+\frac{3}{4}\frac{\delta \dot{\theta}}{\dot{\bar{\theta}}^{1/2}}-\frac{3}{8m}\frac{(\nabla \delta \theta)^2}{\dot{\bar{\theta}}^{1/2}}-\frac{3m}{4k^2}\frac{\delta \rho_b}{\dot{\bar{\theta}}^{1/2}}\right)\right]=-c_2a^3(\bar{\rho}_b+\delta \rho_b)
\end{equation}
The zeroth order E.O.M is given by,
\begin{equation}
\frac{d}{dt}\left[c_1a^3\left(\frac{3}{2}\dot{\bar{\theta}}^{1/2}\right)\right]=-c_2a^3\bar{\rho}_b
\label{zero}
\end{equation}
and the first order E.O.M is,
\begin{equation}
\frac{d}{dt}\left[c_1a^3\left(\frac{3}{4}\frac{\delta \dot{\theta}}{\dot{\bar{\theta}}^{1/2}}-\frac{3m}{4k^2}\frac{\delta \rho_b}{\dot{\bar{\theta}}^{1/2}}\right)\right]=-c_2a^3\delta \rho_b
\label{perturb}
\end{equation}
Now, if we put $\rho_b =0$ in the above two equations, we get $\dot{\bar{\theta}}^{1/2} \propto 1/a^3$ (behaving as dust) and $\delta \dot{\theta}/\dot{\bar{\theta}}^{1/2} \propto 1/a^3$ which implies $\delta_{m}= \frac{\delta \dot{\theta}}{2\dot{\bar{\theta}}}$ is constant in time, which means DM perturbations can not grow if baryons are not present. This might be due to the fact that the superfluid DM particles being very light have negligible gravitational attraction towards each other. On the other hand they have a small but non-zero outward pressure which prevents them from clustering. The particles interacts only through the baryons which allow them to cluster enough for structures to form. 

For obtaining the analytical solution of \eqref{perturb}, we make some simple yet valid assumptions. First of all, let us quote all the parameter values that we use for our calculation. We choose $m= 1$eV, $\alpha=10^{-6}$ and $\Lambda=500$ eV (\cite{Ferreira:2018wup}) which satisfy the constraints imposed on the parameters in the original paper in the cosmological context. In order to solve \eqref{perturb}, we assume the baryon density to evolve as dust i.e. $\rho_b \propto 1/a^3$. Further, we can write the baryon density perturbation (in momentum space) as,
$$\delta \rho_d = \delta_b (k,a) \bar{\rho}_b $$
where $\delta_b (k,a)$ is defined as $P_b(k,a) \propto |\delta_b (k,a)|^2$.\\

At late times ($a \sim 1$), $P_b(k,a)$ falls off as $\sim 1/k^3$ for large $k$ and grows as $k$ for small $k$. The general expression for the power spectrum can be written as: $$ P_b(k,a) \approx f(k) a^2 $$ \\
Using this form in \eqref{perturb}, we get the solution for DM perturbation $\delta_{dm}$ by integrating the R.H.S of \eqref{perturb}. The solution is as follows:
\begin{equation}
\delta \rho_{m} = c_3(k)a-\frac{c_4(k)}{a^{1/2}}+\frac{c_5(k)}{a^3}
\label{dmperturb}
\end{equation} 
where $c_3=4m\delta\rho_{b0} f_1(k)$, $c_4=\frac{3\alpha m^{-3/2}}{5\sqrt{2}M_{Pl}} f_2(k)$, and $c_5(k)$ comes as an integration constant  where the integration is with respect to time. $f_1(k)$ and $f_2(k)$ are general functions of $k$ depending upon which regime we are in. The figure below shows the plot of density perturbation of DM superfluid with respect to the scale factor $a$ as obtained from \eqref{dmperturb}.

\begin{figure}[h!]
\centering
  \includegraphics[width=12cm,height=8cm] {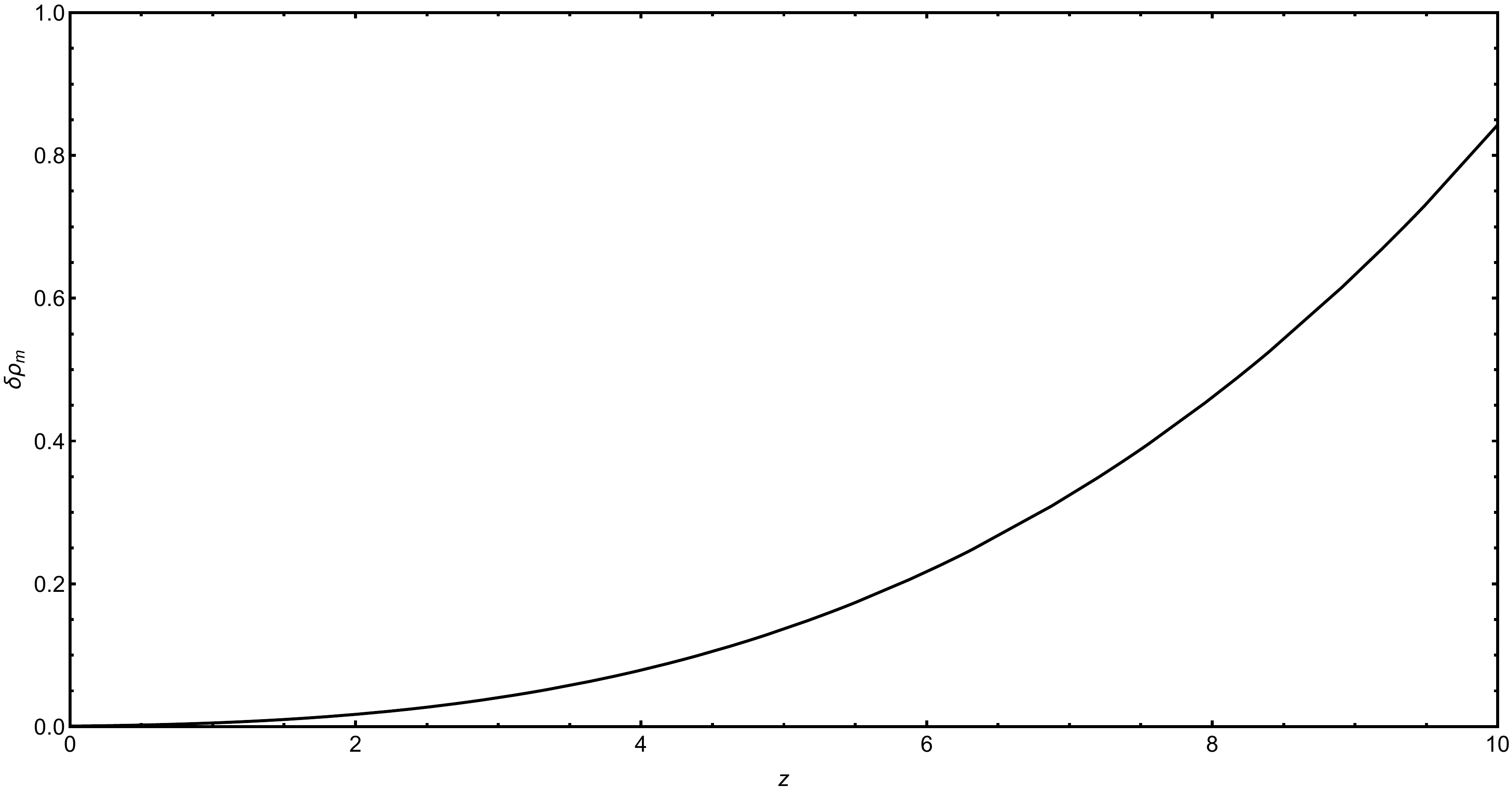}
\caption[Optional caption for list of figures]{Evolution of superfluid dark matter perturbations with $z$} 
\end{figure}

The first term in equation \eqref{dmperturb} suggests that the first order density perturbations grows with time, but as we can see from the plot, this term is not the dominating one.

Now let us look at the other two terms individually. From the second term we have,
$$\delta \rho_{m} \propto \frac{1}{a^{1/2}}$$ Since, for the range of parameter that we consider, the background density evolution behaves like dust ($1/a^3$), this directly gives the relative perturbation growth as,
\begin{equation}
\delta_{m} \propto a^{5/2}
\end{equation}


Important distinct features arise when look at the time evolution of $\delta_{m}$ for each mode. During the matter dominated era, $\delta_{m}$ grows as $a$ in LCDM whereas in this model, it grows as $a^{5/2}$ i.e. at a much faster rate compared to LCDM. For convenience, we write the evolution of $\delta_m$ in terms of the redshift:
\begin{equation}
    \delta_{m} \propto \frac{1}{(1+z)^{5/2}}
\end{equation}

Fig. \ref{new_deltam} below shows the nature of growth in both the models (red solid curve representing LCDM, black dashed curve representing superfluid DM).

\begin{figure}[h!]
\centering
 \includegraphics[width=12cm,height=8cm] {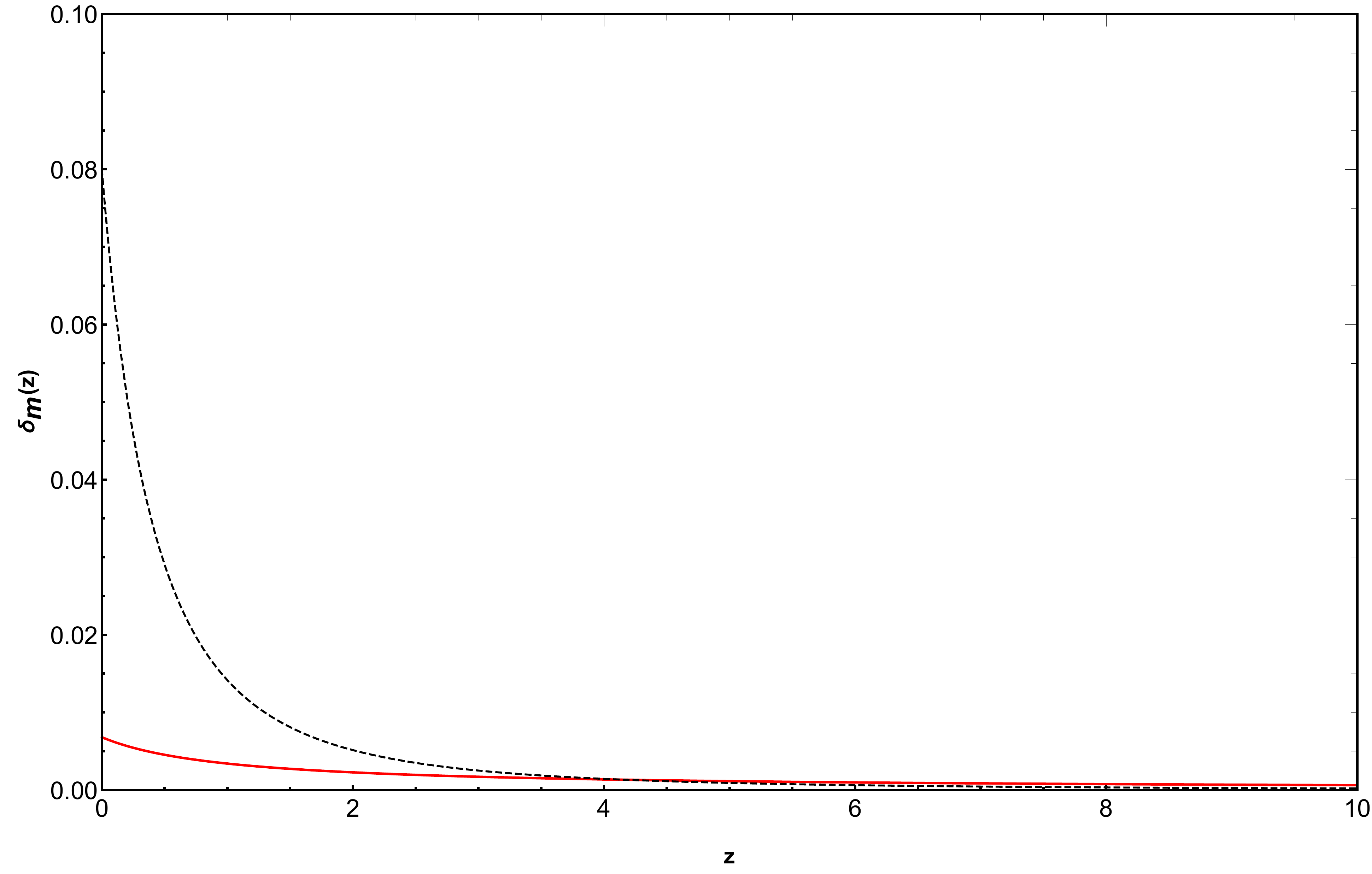}
\caption[Optional caption for list of figures]{Growth of $\delta_{m}$ with respect to $z$. The red solid line represents the growth for $\Lambda$CDM and the black dashed line corresponds to the growth for superfluid. } 
\label{new_deltam}
\end{figure}

Let us now look at the last term of the solution given in \eqref{dmperturb}. The constant $c_3(k)$ appears as an integration constant and is fixed from the observed power spectrum today. From this term we have $$\delta \rho_{m} \propto \frac{1}{a^{3}}.$$ Since this term is proportional to $1/a^3$, this gives a relative density perturbation growth constant in time.

In the next section we solve the perturbation equations numerically in the linear regime and look for any possible deviations from LCDM.

\section{Numerical Solution}

We start with the fluid equations that govern the dynamics of the  dark matter superfluid. The fluid equations, namely the continuity equation and the Navier-Stokes' equation can be derived using the Hamiltonian formalism, as described in \cite{Ferreira:2018wup}. In \cite{Ferreira:2018wup}, the authors work out the fluid equations for an interacting two-component BEC dark matter. Here in this work, we follow the same prescription for a superfluid dark matter which interacts with the baryonic matter.
The corresponding Lagrangian is given by \eqref{lagrangian}.\\

From the Lagrangian, we get the conjugate momentum as,
\begin{eqnarray} \Pi_\theta &=& \frac{\partial \mathcal{L}}{\partial \dot{\theta}} \nonumber \\ &=& \Lambda (2m)^{3/2}\left[\dot{\theta}-m\Phi-\frac{(\nabla \theta)^2}{2m} \right]^{1/2}
\label{momentum}
\end{eqnarray}

The Hamiltonian $H$ describing the superfluid can be obtained as,
\begin{equation}
    H = \Pi_\theta \dot{\theta} - \mathcal{L} 
\end{equation}
Since, $\dot{\theta} = m\Phi +\frac{(\nabla \theta)^2}{2m} +\frac{\Pi_\theta^2}{\Lambda^2(2m)^3}$ from \eqref{momentum},
we get the Hamiltonian $H$ as follows,
\begin{equation}
H= \frac{\Pi_\theta^3}{3\Lambda^2(2m)^3}+\left(m\Phi +\frac{(\nabla \theta)^2}{2m}\right)\Pi_\theta +\frac{\alpha \Lambda}{M_{pl}}\rho_b \theta 
\label{hamiltonian}
 \end{equation}
 \vspace{.3in}
\subsection{Hamilton's equation of motion}

The Hamilton's equations of motion are :
\begin{equation}
    \dot{\theta} = \frac{\partial H}{\partial  \Pi_\theta}
\end{equation}

and

\begin{equation}
    \dot{\Pi}_\theta = -\frac{\partial H}{\partial \theta}
\end{equation}

For this model, the two equations become, respectively,
\begin{equation}
    \dot{\theta} = \frac{\Pi_\theta^2}{\Lambda^2(2m)^3} + m\Phi +\frac{(\nabla \theta)^2}{2m}
    \label{H1}
\end{equation}
and
\begin{equation}
  \dot{\Pi}_\theta = \frac{1}{m} \nabla \cdot (\Pi_\theta \nabla \theta) -\frac{\alpha \Lambda}{M_{pl}}\rho_b 
  \label{H2}
\end{equation}
\vspace{.3in}
\subsection{Fluid equations}
In order to get the fluid equations from the above Hamilton's equations of motion, we identify the terms as corresponding hydrodynamical variables.
 We define the mass density term (as the co-efficient of $\Phi$ in the Hamiltonian) and the four-velocity of the fluid, $\Vec{u}$ as   
\begin{equation} \rho_m = m \Pi_\theta,\quad \Vec{u} = -\frac{\nabla \theta}{m} .  \end{equation}
 Using the above definitions, we get the fluid equations from equation \eqref{H2} and \eqref{H1} as follow, 
 \begin{eqnarray}
     \dot{\rho}_m + \nabla \cdot (\rho_m \Vec{u}) &=& -\frac{\alpha \Lambda m}{M_{pl}}\rho_b \\
    \dot{\Vec{u}} + (\Vec{u} \cdot \nabla) \Vec{u} &=& -\frac{\rho_m \nabla \rho_m}{4 \Lambda^2 m^6} - \nabla \Phi  
 \end{eqnarray}
 
 These are the two fluid equations: Continuity equation and Navier-Stokes' equation.\\
 
 Now, the Poisson's equation can be written as 
 \begin{equation}
     \nabla^2 \Phi = 4 \pi G (\bar{\rho} + \delta \rho)
 \end{equation}
 Integrating twice and substituting the background density using Friedmann equations, we get the potential as: 
  
  \begin{equation}
      \Phi = -\frac{1}{2}(\dot{H} + H^2)l^2 + \phi
  \end{equation}
  where $l$ is the proper distance defined as $\Vec{l} = a(t) \Vec{x}$ and $\phi$ is the potential due to inhomogeneities.  \\
  Similarly, the four-velocity $\Vec{u}$ can be split into two parts, Hubble flow and a peculiar velocity $\Vec{v}$ as follows:
  
  \begin{equation}
      \Vec{u} = H \Vec{l} + \Vec{v}
  \end{equation}
  
  Expressing everything in comoving co-ordinates $\Vec{x}$ and using $\nabla_l = \frac{1}{a(t)}\nabla_x$, we get,
  
  \begin{equation}
    \dot{\rho}_m + 3H\rho_m + \frac1a \nabla \cdot (\rho_m \Vec{v}) = -\frac{\alpha \Lambda m}{M_{pl}}\rho_b  
  \end{equation}
  
  and 
  \begin{equation}
      \dot{\Vec{v}} + H\Vec{v}+ \frac1a (\Vec{v} \cdot \nabla) \Vec{v} = -\frac{\rho_m \nabla \rho_m}{4a \Lambda^2 m^6} - \frac{\nabla \phi}{a} 
  \end{equation}
  
  These are the two fluid equations of motion that we shall use for the rest of our calculations.\\
  \vspace{.3in}
  \subsection{Perturbations}
  The total DM density $\rho_m$ and the baryonic density $\rho_b$ can be split into two parts: background and perturbation:\\
  $$ \rho_m = \bar{\rho}_m + \delta\rho_m, \quad  \rho_b = \bar{\rho}_b + \delta\rho_b$$ 
  respectively.\\
  We define, the relative density perturbations for these two components as,\\
  $$ \delta_m = \frac{\delta\rho_m}{\bar{\rho}_m + \bar{\rho}_b} \quad {\rm and} \quad \delta_b = \frac{\delta\rho_b}{\bar{\rho}_m + \bar{\rho}_b}.$$
  In the linear perturbation regime, we treat $\delta\rho_m$, $\delta\rho_b$ and $\Vec{v}$ to be small, and hence, neglect the higher orders of these terms. Perturbing the two fluid equations in the linear regime gives:
  \begin{equation}
      \dot{\delta}_m + \frac{\bar{\rho}_m}{a \bar{\rho}}\nabla \cdot\Vec{v} = -\frac{\alpha \Lambda m}{M_{pl}}\delta_b
      \label{deltam}
  \end{equation}
  and
  \begin{equation}
      \dot{\Vec{v}} + H \Vec{v} = -\frac{\bar{\rho}_m \nabla \delta \rho_m}{4a \Lambda^2 m^6} - \frac{1}{a} \nabla \phi
      \label{vel}
  \end{equation}
  By using the above equations along with the Poisson's equation, we get the evolution equation for $\delta_m$ as follows:
\begin{equation}
    \Ddot{\delta_m}+\frac{a^2 \bar{\rho}_m \delta_m}{2M_{pl}^2} -\frac{\bar{\rho}_m^2 \nabla^2 \delta_m}{4\Lambda^2 m^6}= H \nabla \cdot \Vec{v} -\frac{\alpha \Lambda m \delta_b}{M_{pl}}-\frac{a^2 \bar{\rho} \delta_b}{2 M_{pl}^2}
\end{equation}
This is a second order differential equation. The coefficient of the spatial derivative $\nabla^2$ gives the square of the sound speed $c_s$. Thus, we get,
\begin{equation}
    c_s^2 = \frac{\bar{\rho}_m^2}{4\Lambda^2 m^6}
\end{equation}
Below in Fig. \ref{cs}, we show the plot for $c_s^2$ vs the redshift $z$ for $m = 1$ eV and $\Lambda = 500$ eV. We take the time evolution of the background density $\bar{\rho}_m$ as 
\begin{equation}
    \bar{\rho}_m = \frac{0.4(1+z)^3}{(1+1000)^3}
    \label{background}
\end{equation}
where the value of $\bar{\rho}_m$ at equality ($z=1000$) is set as $0.4$ eV$^4$ (\cite{Ferreira:2018wup}).
\begin{figure}[h!]
\centering
 \includegraphics[width=12cm,height=8cm] {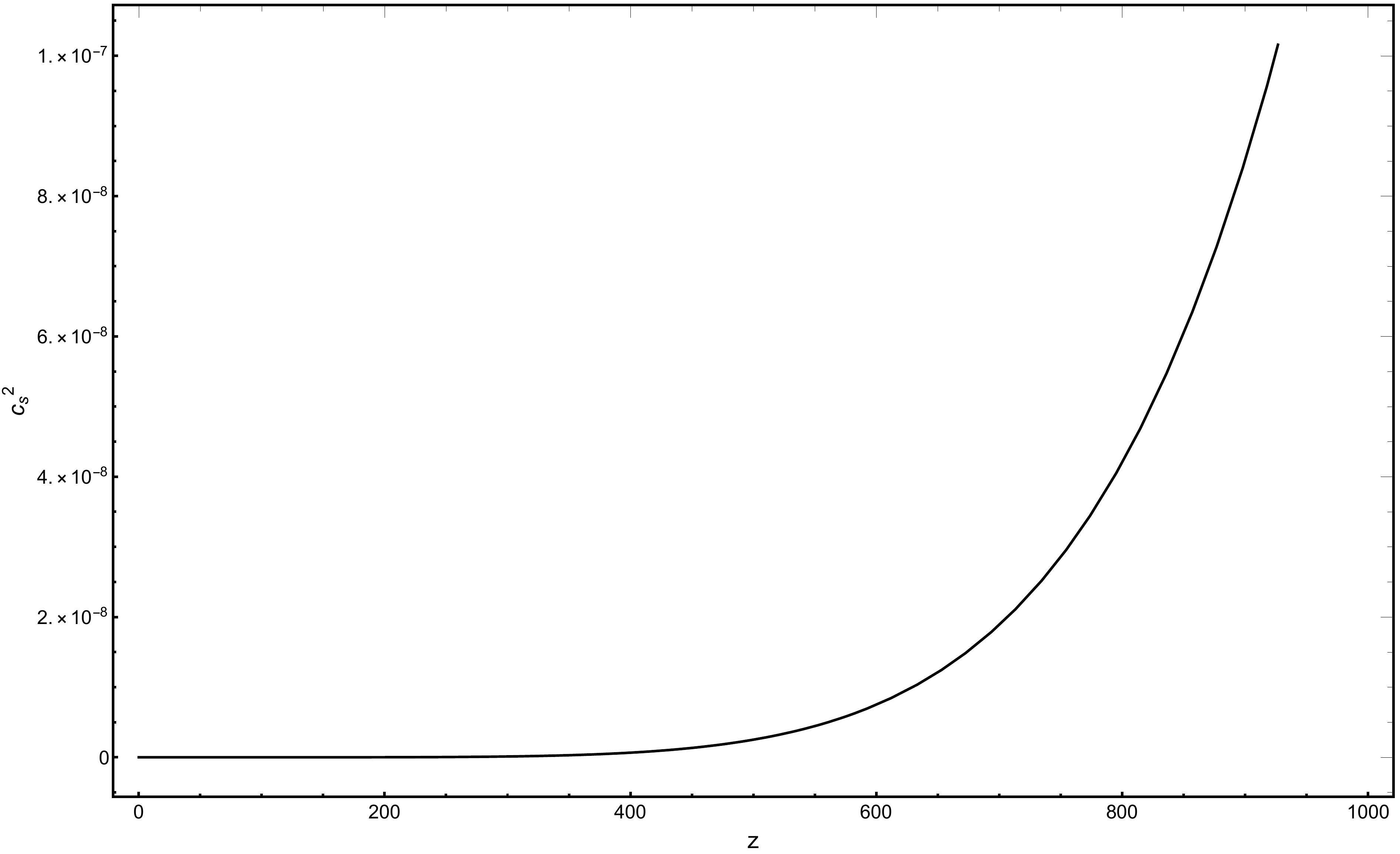}
\caption[Optional caption for list of figures]{Plot for $c_s^2$ vs $z$} 
\label{cs}
\end{figure}
As evident from the plot, the sound speed is very small (compared to the speed of light $c=1$). 
\vspace{.5in}

In order to obtain the solutions for $\delta_m$, we rewrite
equations \eqref{deltam} and \eqref{vel} in the Fourier domain in physical co-ordinate as,
\begin{equation}
      \dot{\delta}_m + \frac{\bar{\rho}_m}{\bar{\rho}}(ikv) = -\frac{\alpha \Lambda m}{M_{pl}}\delta_b
  \end{equation}
  and
  \begin{equation}
      ik\dot{v} + ik H v = \frac{k^2\bar{\rho}_m \bar{\rho} \delta_m}{4 \Lambda^2 m^6} + \frac{a^2}{2M_{pl}^2}\left(\delta \rho_m+\delta \rho_b + \frac{3 i a H \bar{\rho} v}{k}\right)
  \end{equation}
  
  To solve the above equations, we write them in terms of redshift,
  \begin{equation}
      -H(1+z)\frac{d\delta_m}{dz} + \frac{\bar{\rho}_m}{ \bar{\rho}}(ikv) = -\frac{\alpha \Lambda m}{M_{pl}}\delta_b
  \end{equation}
  and
  \begin{equation}
      -ikH(1+z)\frac{dv}{dz} + ik H v = \frac{k^2\bar{\rho}_m \bar{\rho}\delta_m}{4 \Lambda^2 m^6} +  \frac{ \bar{\rho}}{2M_{pl}^2(1+z)^2}\left(\delta_m + \delta_b + \frac{3 i H v}{k (1+z)}\right)
  \end{equation}
  \vspace{.3in}
  \textbf{Parameters and initial conditions:}\\
  The model parameters involved are $m$, $\Lambda$ and $\alpha$. We take $m= 1$ eV and $\Lambda= 500$ eV while keeping the parameter $\alpha$ as free parameter which is varied to check where the model deviates from flat LCDM.\\
  
  We integrate the perturbation equations using the following initial conditions at the epoch of equality $z = 1000$:
  We set $\delta_b(z =1000) = \delta_m(z=1000) = 10^{-5}$ and $H(z=1000)= m = 1$ eV.\\
  Since $\bar{\rho_m} \gg \bar{\rho}_b$, we assume $\bar{\rho} = \bar{\rho}_m +\bar{\rho}_b \approx \bar{\rho}_m$ as given in \eqref{background}.\\
  The initial value of $v$ at $z=1000$ is chosen to be around $1$. For the time evolution of the background density and Hubble parameter, we take the usual LCDM evolution of these quantities in matter-dominated era, i.e. $\bar{\rho}_m \propto 1/a^3$ and $H \propto 1/a^{3/2}$. Furthermore, we take $\delta_b \propto a$. We keep the wavenumber $k$ fixed at $0.0001$ eV, although the nature remains same for larger values of $k$. 
  
\begin{figure}[h!]
\centering
 \includegraphics[width=12cm,height=8cm] {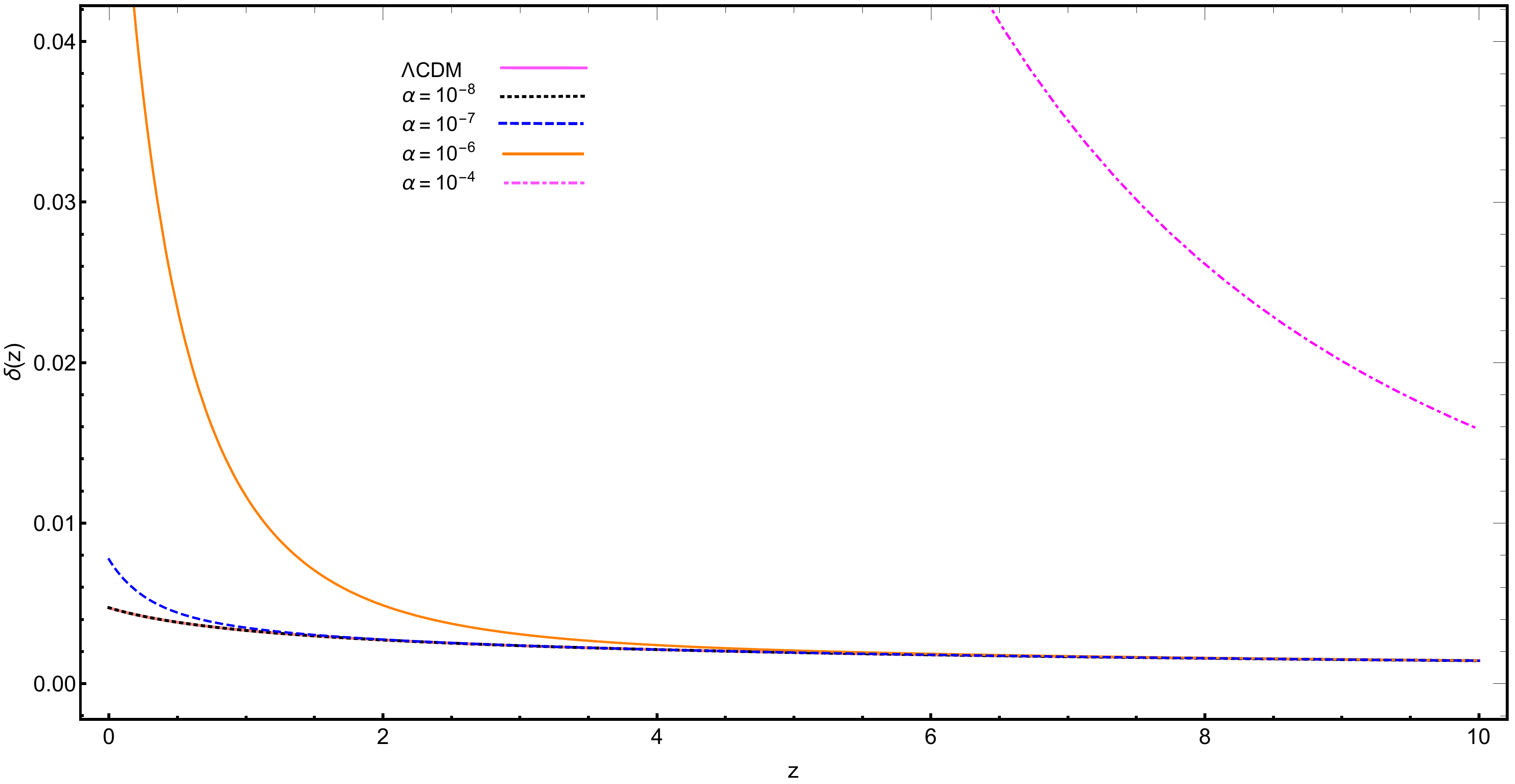}
\caption[Optional caption for list of figures]{$\delta_m$ vs $z$} 
\label{plot2}
\end{figure}

 Figure \ref{plot2} shows the evolution of the DM density perturbation $\delta_m$ with respect to the redshift $z$ for different values of $\alpha= 10^{-8},\ 10^{-7}, \ 10^{-6},\ 10^{-4} $ and also for LCDM corresponding to $\alpha = 0, \Lambda \to \infty$. As expected, the smaller the value of $\alpha$, the closer the resemblance with LCDM-like evolution. As we see in Fig. \ref{plot2}, the plot for $\alpha=10^{-8}$ coincides with LCDM. When $\alpha$ is large enough, the growth is very steep. This is because a large enough $\alpha$ implies large interaction strength between the superfluid phonons and baryons, ensuring that structure formation takes place at an earlier epoch as compared to LCDM. 
  

\begin{figure}[h!]
\centering
 \includegraphics[width=12cm,height=8cm] {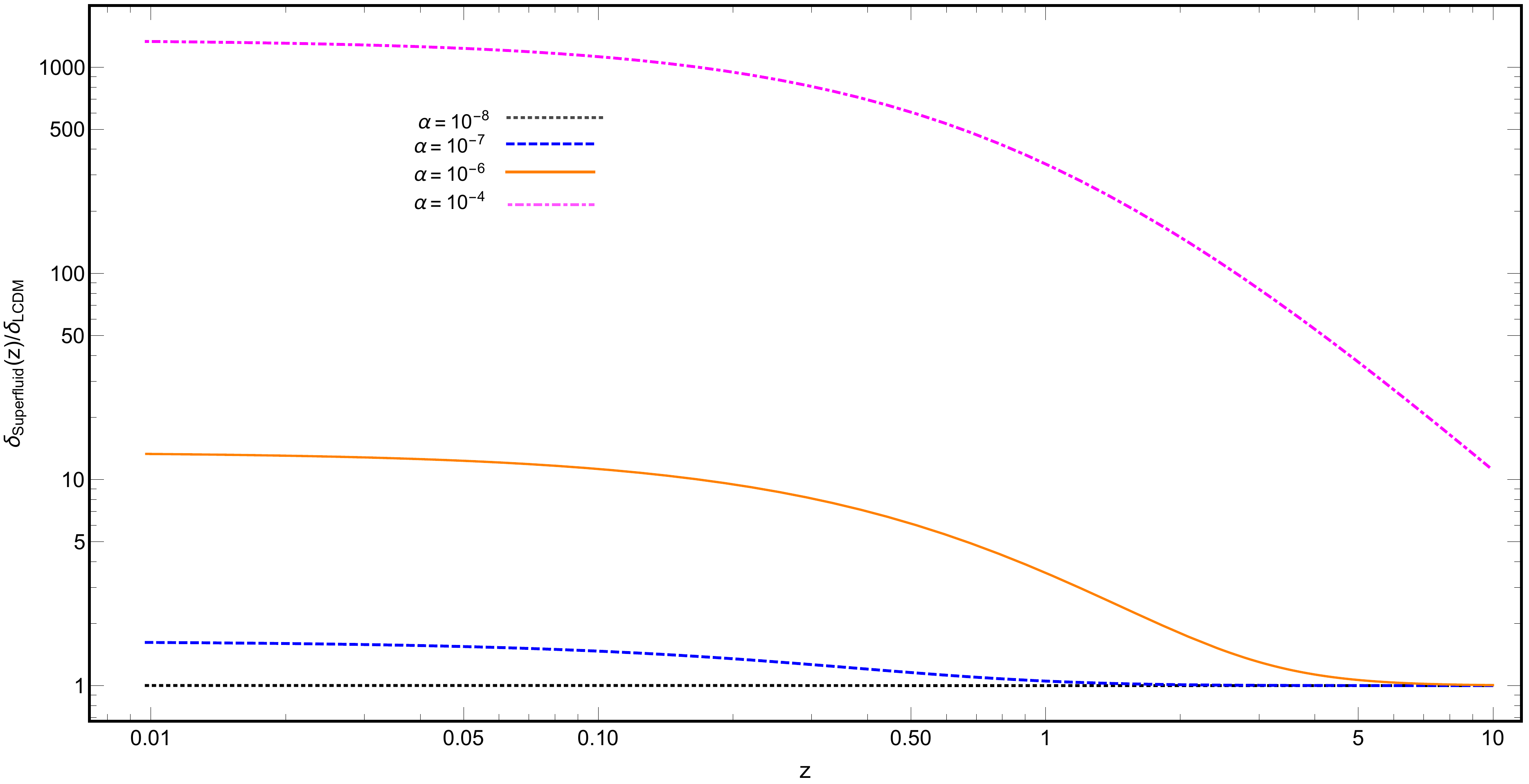}
\caption[Optional caption for list of figures]{$\delta_{{\rm superfluid}}/\delta_{LCDM}$ vs $z$} 
\label{plot3}
\end{figure}

In figure \ref{plot3},  we plot the relative differences between the perturbation growth in LCDM model and superfluid DM model for different values of $\alpha$ in terms of $\delta_{{\rm superfluid}}/\delta_{LCDM}$. As expected, the ratio is very high at a lower redshift. As we go to higher redshifts, the ratio tends to $1$ i.e., they eventually agree with LCDM  at very high redshifts and matches exactly at $z=1000$ where we set our initial conditions. The LCDM model corresponds to $\alpha =0$. For $\alpha = 10^{-8}$, the deviation from LCDM at low redshift goes upto $0.13\%$ at $z= 0.01$. The larger the value of $\alpha$, the higher is the ratio, implying a stronger deviation from LCDM at low enough redshifts. As $\alpha$ is increased to $10^{-7}$, the deviation from LCDM becomes much larger ($ \sim 62\%$). 
\begin{figure}[h!]
\centering
 \includegraphics[width=12cm,height=8cm] {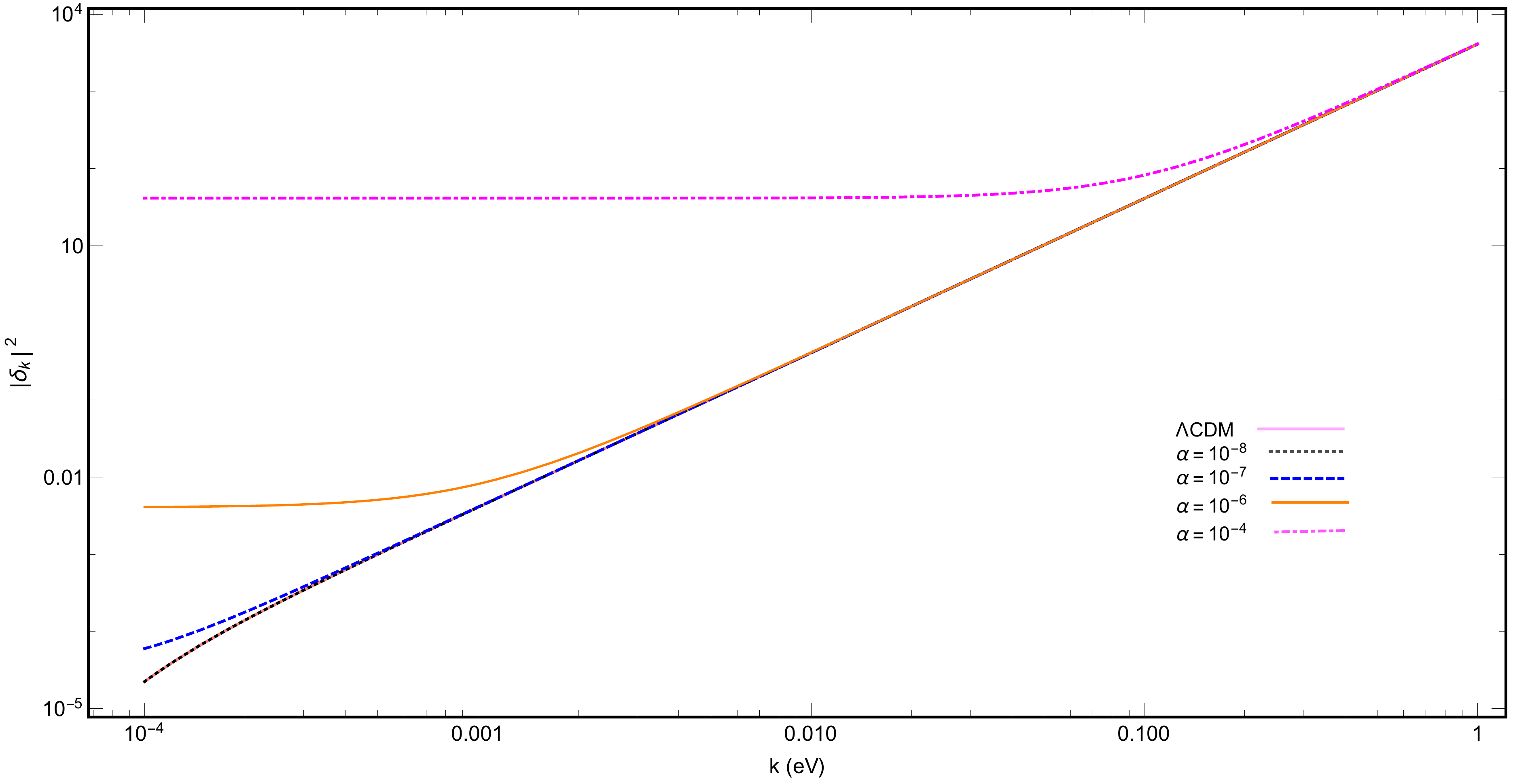}
\caption[Optional caption for list of figures]{$|\delta_k|^2$ vs $k$} 
\label{power}
\end{figure}
We can also plot the matter power spectrum $P(k)$ as a function of $k$ at $z=0$. The matter power spectrum $P(k) \propto |\delta_m(k)|^2$. In Fig. \ref{power}, we plot $|\delta_m(k)|^2$ vs. $k$ which shows how the power varies for different values of $\alpha$. As shown in the figure, the power spectrum for $\alpha = 10^{-8}$  matches with the LCDM prediction. As can be seen, the power increases for larger values of $\alpha$ at a given value of $k$. This is because the perturbation growth is stronger for large $\alpha$ as discussed earlier.

\section{Results and Discussions}

The superfluid dark matter model is a very promising and newly emerging model of cosmology combining together the rich physics of condensed matter, particle physics and cosmology. In view of its success in explaining a number of observations within the galaxies where LCDM fails to provide a satisfactory explanation, this model can be said to offer a greater understanding of the universe. In their earlier works, Khoury and his collaborators have investigated the implications of this model at galactic scales. However, a complete study of cosmological implications have not been performed earlier. In this paper, we have tried to investigate, both analytically and numerically, whether the predicted cosmology of the model tallies well with the observations and how different the predictions are from that of LCDM. In the realm of non-relativistic low energy effective theory of superfluid, the background cosmology agrees with the predictions of LCDM, and this gives a constraint on the two model parameters $\alpha$ and $\Lambda$ which turn out to be different than their galactic scale constraints. This result has also been discussed in \cite{khoury2015}. At the level of first order perturbation, we find that the above constraints lead to a cosmology which differ significantly from LCDM. In particular, our analytical results suggests that the growth of density perturbations of dark matter superfluid roughly goes as $a^{5/2}$, which is much higher compared to the LCDM picture ($\delta_{dm} \propto a$). This might be due to the strong interaction between superfluid phonons and baryonic matter. This behaviour has also been verified from the numerical solutions. For the numerical calculations, in particular, we have kept two of the model parameters $m$ and $\Lambda$ fixed at $1$ eV and $500$ eV respectively. This gives an upper bound on the third parameter: $\alpha \leq 10^{-8}$ corresponding to just $0.13\%$ deviation from LCDM. This is different from the value quoted in \cite{khoury2015}. The bound obtained in \cite{khoury2015}, for $m=1$ eV, is $\alpha \leq 10^{-4}$, which, even though predicts the correct background evolution, strongly deviates from LCDM in the context of perturbation growth in the present epoch. This can be seen in Figs. \ref{plot2}, \ref{plot3} and \ref{power}. In our analysis, we have assumed the baryonic component to follow standard dust evolution ($\propto \frac{1}{a^3}$). 

We also notice from \eqref{zero} and \eqref{perturb} that the superfluid dark matter perturbations do not grow in the absence of baryons. We think that this is due to the extremely small mass of these dark matter particles. Being very light, they have negligible gravitational pull towards each other. They also have a certain non-zero outward pressure due to their superfluid nature. Hence, this prevents them from clumping together significantly. Only when baryons are present, they can interact strongly enough through a MONDian force and can form structures.   

A more complete analysis of the perturbation growth should rely on the proper relativistic extension of the theory, which has not been attempted in this paper. Some relativistic models have been discussed in the original paper \cite{khoury2015}, however a rigorous analysis is still lacking. We hope to address the same in a future work. Our work looks into the solution in the linear regime where perturbations are taken to be small. In future, we plan to extend our analysis to the non-linear regime and study the structure formation through spherical collapse. It would also be interesting to see how well this model predicts the CMB or the halo mass function.



\acknowledgments

S. Banerjee would like acknowledge funding from Israel Science Foundation and John and Robert Arnow Chair of Theoretical Astrophysics. S. Bera would like to acknowledge support from Navajbai Ratan Tata Trust and IUCAA research grant. S. Banerjee and S. Bera would also like to thank the hospitality of the Tata Institute of Fundamental Research, India, where part of the work was done. DFM thank the Research Council of Norway for their support. Computations were performed on resources provided by UNINETT Sigma2 -- the National Infrastructure for High Performance Computing and Data Storage in Norway.



\end{document}